\documentclass{tlp}

\usepackage[utf8]{inputenc}
\usepackage{latexsym}
\usepackage{graphicx}
\usepackage{subfigure}

\title{The YAP Prolog System}

\author[V.\ Santos Costa et al.]
  { V\'{I}TOR SANTOS COSTA and RICARDO ROCHA\\
  DCC \& CRACS INESC-Porto LA, Faculty of Sciences, University of Porto\\
  R. do Campo Alegre 1021/1055, 4169-007 Porto, Portugal\\
  \email{\{vsc,ricroc\}@dcc.fc.up.pt}\\
  \and LU\'{I}S DAMAS\\
  LIACC, Faculty of Sciences, University of Porto\\
  R. do Campo Alegre 1021/1055, 4169-007 Porto, Portugal\\
  \email{luis@ncc.up.pt}\\ }

\submitted {10 October 2009}
\revised{5 March 2010}
\accepted{6 February 2011}

\begin{document}
\bibliographystyle{acmtrans}

\long\def\comment#1{}

\maketitle

\setcounter{page}{1}

\maketitle

\setcounter{footnote}{0} \begin{abstract} 

  Yet Another Prolog (YAP) is a Prolog system originally developed in
  the mid-eighties and that has been under almost constant development
  since then. This paper presents the general structure and design of
  the YAP system, focusing on three important contributions to the
  Logic Programming community. First, it describes the main techniques
  used in YAP to achieve an efficient Prolog engine. Second, most
  Logic Programming systems have a rather limited indexing
  algorithm. YAP contributes to this area by providing a dynamic
  indexing mechanism, or just-in-time indexer (JITI). Third, a
  important contribution of the YAP system has been the integration of
  both or-parallelism and tabling in a single Logic Programming
  system.

\end{abstract}


\section{Introduction}
\label{sec:indexation}


Prolog is a widely used Logic Programming language. Applications
include the semantic web~\cite{DBLP:conf/w3c/DevittRC05}, natural
language analysis~\cite{1137806},
bioinformatics~\cite{DBLP:conf/iclp/Mungall09}, machine
learning~\cite{DBLP:journals/jmlr/PageS03}, and program
analysis~\cite{DBLP:conf/ppdp/BentonF07}, just to mention a
few. 
In this paper, we discuss the design of the \emph{Yet Another Prolog
  (YAP)} system and discuss how this system tries to address the
challenges facing modern Prolog implementations.
First, we present the general structure and organization of the system
and then we focus on three contributions of the system to the Logic
Programming community: \emph{engine design}, the \emph{just in-time
  indexer}, and \emph{parallel tabling}.  Regarding the first
contribution, one major concern in YAP has always been to maintain an
efficient interpreted Prolog engine. The first implementation of the
YAP engine achieved good performance by using an emulator coded in
assembly language. Unfortunately, supporting a large base of assembly
code raised a number of difficult portability and maintenance
issues. Therefore, more recent versions of YAP use an emulator written
in \texttt{C}. A significant contribution of our work was to propose a
number of techniques for Prolog emulation and show that these
techniques can lead to significant increases in
performance~\cite{yap-optim}. Although, our initial concern was
execution speed, memory usage is also a significant issue in several
Prolog applications, namely if the applications manipulate large
databases. YAP implements a number of techniques to reduce total
memory usage in this case~\cite{DBLP:conf/padl/Costa07}.

Ideally, Logic Programming should be about specifying the logic of the
program, and then provide control. In practice, Logic Programming
systems can often be very vulnerable to seemingly irrelevant details
such as argument order. Especially for larger databases, swapping
order of arguments may result in order of magnitude speed improvements
for some programs. As such databases become more common, these
problems become more important. YAP contributes to this area by
providing a dynamic indexing mechanism, or just-in-time indexer
(JITI)~\cite{jit-index}. The JITI alleviates questions of argument
order, as it can make Prolog competitive in applications that would
otherwise require a database manager~\cite{icia05}.

A third contribution of the YAP system has been the integration of
\emph{both} or-parallelism and tabling in a single Logic Programming
system. Inspired by previous research on the Muse
system~\cite{Ali-90b} and on the XSB engine~\cite{Sagonas-98}, YAP was
the first engine to actually integrate these two very different, and
yet related, mechanisms into a single engine, OPTYAP~\cite{optyap-journal}.
In our experience, the YAP tabling mechanisms are the most widely used
extension of YAP, and are a key focus for the future of our system.
Parallelism has been a less widely used feature of YAP, although our
work in supporting parallelism was most beneficial in implementing the
YAP \emph{threads} library. Recent advances in computer architecture
have rekindled interest in implicit parallelism in
YAP~\cite{DBLP:journals/tplp/CostaDR10}.


The paper is organized as follows. In Section~\ref{sec:history}, we
first give a brief overview of the system history, adapted
from~\cite{DBLP:conf/iclp/Costa08}. Next, in
Section~\ref{sec:org} we present the general structure and we
discuss the main data-structures in the YAP
engine. Section~\ref{sec:engine} presents the main contributions in
the engine, and Section~\ref{sec:compiler} discusses the design of the
compiler. The two are tightly
integrated~\cite{yap-optim}. Section~\ref{sec:jiti} discusses the
implementation of the JITI, and Section~\ref{sec:optyap} presents
OPTYAP. We conclude by discussing some of the main issues in our work
in Section~\ref{sec:challenges}, and present conclusions in
Section~\ref{sec:conclusions}. Throughout the text, we assume the
reader will have good familiarity with the general principles of
Prolog implementation, and namely with the WAM~\cite{Warren83}.

\section{A Little Bit of History}
\label{sec:history}

The history of Prolog and Logic Programming starts in the early
seventies, with the seminal works by Colmerauer, Roussel, and
Kowalski~\cite{Col93}. The original Marseille Prolog was promptly
followed by quick progress in the design of Logic Programming systems.
One of the most exciting developments was David H.\ D.\ Warren's
abstract interpreter, eventually called the Warren Abstract Machine or
WAM~\cite{Warren83}, which became the foundation of Quintus
Prolog~\cite{quintus}. The success of Quintus Prolog motivated the
development of several Prolog systems. Yet Another Prolog (YAP) is one
example, and was started by Luís Damas and colleagues in 1984 at the
University of Porto. Luís Damas had returned from the University of
Edinburgh, where he had completed his PhD on type
systems~\cite{DBLP:conf/popl/DamasM82}. He was also interested in
Logic Programming and, while at Edinburgh, had designed one of the
first Prolog interpreters, written in the IMP programming language for
the EMAS operating system, which would become the basis for the famous
C-Prolog interpreter~\cite{c-prolog}. Together with Miguel
Filgueiras, who also had experience in Prolog
implementation~\cite{DBLP:books/eh/campbell84/Filgueiras84}, they
started work on the development of a new WAM-based Prolog. The goal
was to design a compact, very fast system emulator, written in
assembly. To do so, Luís Damas wrote the compiler in \texttt{C} and an
emulator in \texttt{68000} assembly code.

Arguably, one of the strengths of YAP derives from Luís Damas'
experience in Edinburgh: internal object representation was well defined
from the start and always facilitated development. YAP included
several improvements over the original WAM design: it used a
depth-first design to visit terms, and it was one of the first Prologs
to do indexing on sub-terms~\cite{costa88}. YAP also provided a very
fast development environment, due to its \texttt{C}-written
compiler. The combination of fast compilation and execution speed
attracted a dedicated user community, mainly in Artificial Intelligence
(e.g., Moniz Pereira's group supported the first YAP port to the VAX
architecture). A major user was the European Union Eurotra
project~\cite{eurotra} for which YAP developed \emph{sparse functors}:
one of the first attempts at using named fields for structures in
Prolog.

The second chapter in YAP's history started on the mid-nineties. At
this point in time, YAP development had slowed down. One problem was
that the system had become very complex, mainly due to the need to
support several instruction set architectures in assembly (at the
time: 68000, VAX, MIPS, SPARC, HP-RISC, Intel x86). Unfortunately, a
first attempt at using a \texttt{C} interpreter resulted in a much
slower system. On the other hand, the user community was not only
alive but growing, as Rui Camacho had taken YAP to the Turing
Institute Machine Learning Group, where it was eventually
adopted by Inductive Logic Programming (ILP) systems such as P-Progol,
later Aleph~\cite{aleph-manual}, and IndLog~\cite{indlog}. Second,
researchers such as Vítor Santos Costa and Fernando Silva, had
returned to Porto and were interested in Parallel Logic Programming.
While SICStus Prolog would have been an ideal platform, it was a
closed source system. YAP therefore became a vehicle of research first
in parallelism~\cite{Rocha-99b} and later in
tabling~\cite{optyap-journal}. A new, fast, \texttt{C}-based emulator was
written to support this purpose~\cite{yap-optim} and brought YAP back
to the list of the fastest Prolog systems~\cite{demoen00}.

Interest in YAP grew during the late nineties, leading to the third
chapter in YAP's story. As hardware scaled up and users had more data
to process, limitations in the YAP design become clear: Prolog
programs perform well for small applications, but often just crash or
perform unbearably slowly as application size grows. Effort has
therefore been invested in rethinking the basics, step by step. The
first step was rewriting the garbage collector~\cite{gc-iclp01}. But
the main developments so far have been in indexing: it had become
clear that the WAM's approach to indexing simply does not work for
applications that need to manipulate complex, large,
databases. Just-In-Time indexing~\cite{jit-index} tries to address
this problem.

\section{System Organization}
\label{sec:org}

\begin{figure}[htb]
  \begin{center}
   \includegraphics[width=12cm]{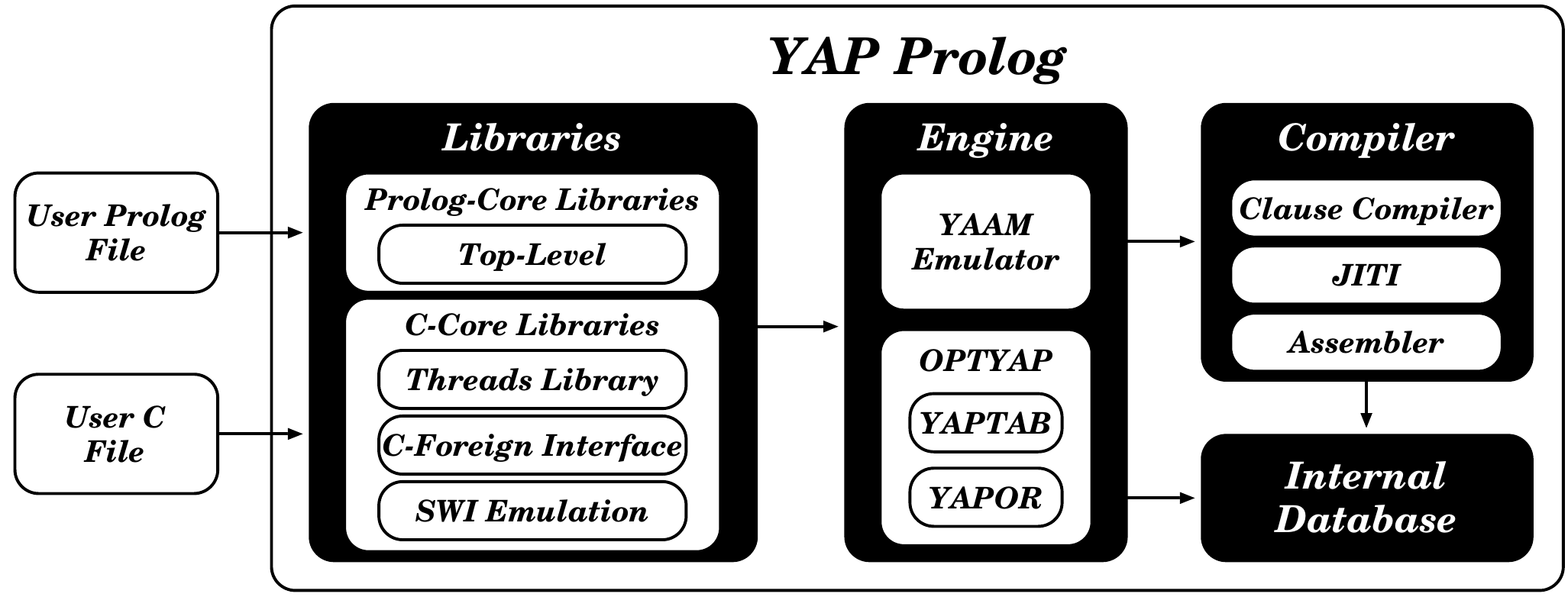}
    \caption{The Organization of the YAP Prolog system. At heart of
      the system we show the YAAM emulator with the OPTYAP extensions.
    The engine is supported by a core-set of libraries written in
    \texttt{C}. These libraries can be user-extended
    through YAP's native foreign language interface and through SWI's
    interface. The compiler and JITI mechanisms are controlled by the
    engine and generate code to be stored in the internal database.}
    \label{fig:structure}
 \end{center}
\end{figure}

Figure~\ref{fig:structure} presents a high-level view of the YAP
Prolog system. The system is written in \texttt{C} and Prolog.
Interaction with the system always starts through the top-level Prolog
library. Eventually, the top-level refers to the core
\texttt{C} libraries. The main functionality of the core \texttt{C}
libraries includes starting the Prolog engine, calling the
Prolog clause compiler, and maintaining the Prolog internal
database. The Prolog sequential engine executes YAP's YAAM
instructions~\cite{yap-optim}, and has been extended to support
tabling and or-parallelism~\cite{optyap-journal}.  The engine may also
call the just-in-time indexer (JITI)~\cite{jit-index}. Both the
compiler and the JITI rely on an assembler to generate code
that is stored in the internal database.

The \texttt{C}-core libraries further include the parser and several
built-ins (not shown in Figure~\ref{fig:structure}). An SWI-compatible
threads library~\cite{SWI-threads} provides support to thread creation
and termination, and access to locking. The Foreign Language Interface
(FLI) library allows external \texttt{C}-code to use the YAP
data-structures. YAP also provides an SWI FLI emulator that translates
SWI-Prolog's~\cite{SWI} FLI to YAP FLI calls. SWI-Prolog packages such
as \texttt{chr}~\cite{DBLP:conf/iclp/Schrijvers08}, \texttt{JPL},
\texttt{SGML}~\cite{SWI}, and even the core SWI-Prolog Input/Output
routines (\texttt{PLStream}) have been adapted to use this layer.

\subsection{The Key Data-Structures}
\label{sec:data_structures}

Throughout, the YAP implementation uses \emph{abstract types} to refer
to objects with similar properties, say, the type \texttt{Term} refers
to all term objects. Each abstract type may have different
concrete types (or subtypes), but a concrete type has a single
abstract type by default. For example, the abstract type \texttt{Term}
has the concrete type \texttt{Appl} (compound term), \texttt{Pair}
(lists), or \texttt{Int}. In all cases, given a subtype \texttt{T} of
some abstract type \texttt{A} the following three functions should be
available:

\begin{itemize}
\item Given a concrete object of concrete type \texttt{T}, the
  $Abs\mathtt{T}$ routine returns an instance of its abstract type
  \texttt{A}.
\item Given an instance of  the abstract type \texttt{A}, the
  $Rep\mathtt{T}$ routine returns an object with concrete type
  \texttt{T}. 
\item Given some arguments, the constructor $Mk\mathtt{T}\mathtt{A}$
  constructs an object of concrete type \texttt{T}, and returns an
  instance of \texttt{A}.
\end{itemize}

For example, given a pointer to the stack, the function $AbsAppl$
returns a \texttt{Term} object; given an object of type \texttt{Term},
the function $RepAppl$ returns a pointer to the stack; and, lastly,
the function $MkApplTerm$ receives a functor, an arity, and an array
of terms and returns an object of type \texttt{Term}. In order to
achieve efficiency, most of these functions are implemented as inline
\texttt{C}-functions. 

\subsection{The Database}

YAP includes two main data-structures, the Engine Context and
the Database. The \emph{Engine Context} maintains the abstract machine
internal state, such as abstract registers, stack pointers, and active
exceptions.  The \emph{Database} data structure maintains the root
pointers to the internal database, including the \emph{Atom Table} and
the \emph{Predicate Table}. The table is accessible from a root
pointer so that the state of the engine can be \emph{saved} to and
\emph{restored} from a dump file.

In order to support parallelism and threads, YAP organizes the
database as:
\begin{itemize}
\item The \texttt{GLOBAL} structure, that is available to all workers;
  locks should protect access to these data-structures.
\item An array of per-worker structures, where each one is called
  \texttt{LOCAL}. We define a worker to be a scheduling unit that can
  run an YAAM engine, that is, a thread or a parallel process. The
  engine abstract registers are accessible through the worker's
  \texttt{LOCAL}.
\end{itemize}

\begin{figure}[htb]
  \begin{center}
   \includegraphics[width=12cm]{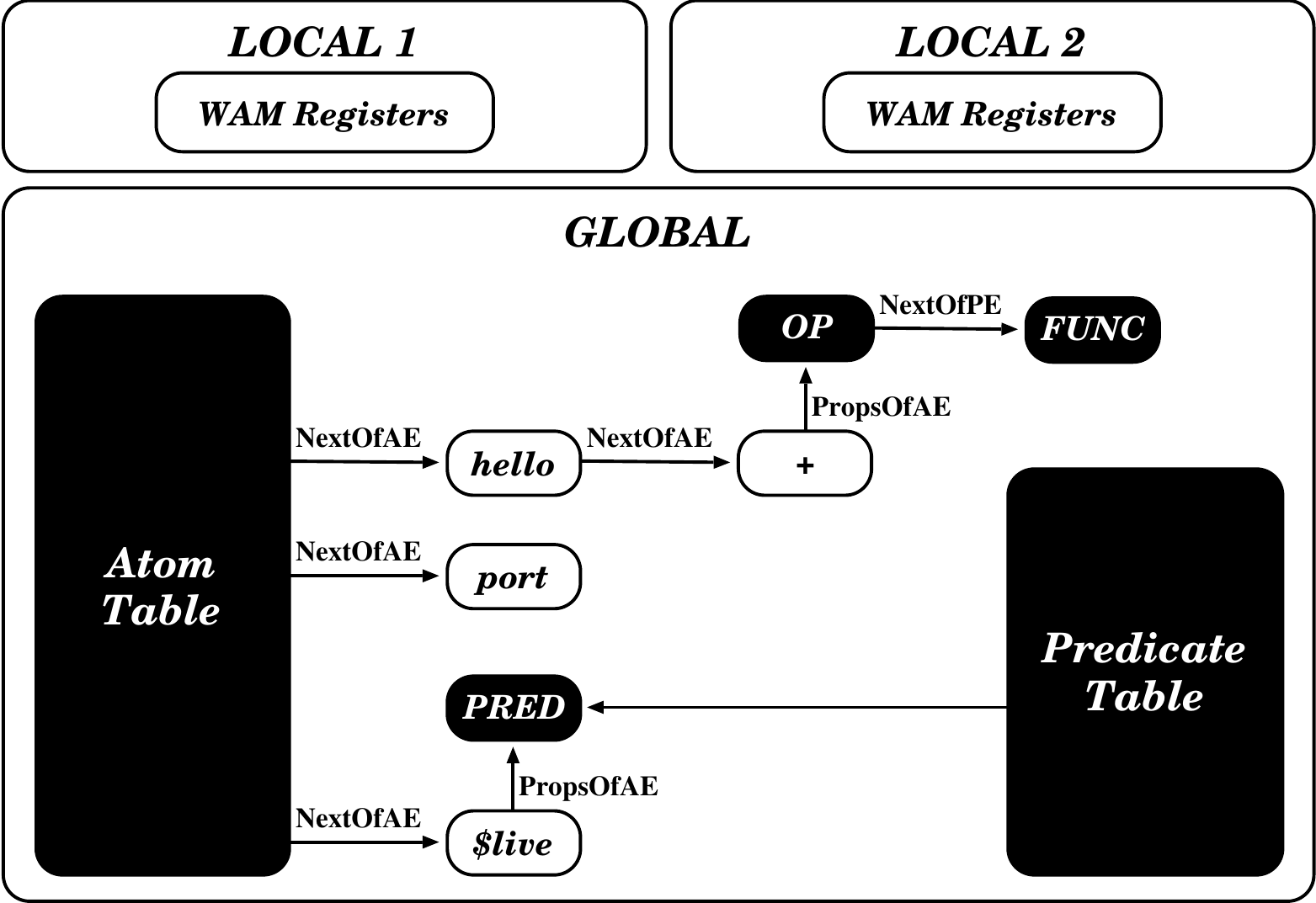}
    \caption{Organization of the YAP database, Each
      worker has a local set of variables (including abstract machine
      registers). All workers share a \texttt{GLOBAL} data structure that
      includes a hash-based atom table. Atoms are stored in a
      linked-list, and most of their properties, including predicates
      with the atom as name, are stored in a linked list for each
      atom. Predicates are often used, so there is a direct hash-table
      for them.}
    \label{fig:atom_table}
 \end{center}
\end{figure}
 
The structure of the database is presented in
Figure~\ref{fig:atom_table}.  We assume support for two workers, hence
we require two \texttt{LOCAL}. Notice that each \texttt{LOCAL}
structure contains a copy of the YAAM registers. The main structure in
the YAP database is the \emph{Atom Table}, containing objects of
abstract type \texttt{Atom}.
The abstract type \texttt{Atom} has a single concrete type,
\texttt{AtomEntry}. Thus, the Atom Table is implemented as a hash table
with linked lists of \texttt{AtomEntry} objects. Each
\texttt{AtomEntry} contains

\begin{enumerate}
\item \texttt{NextOfAE}: a pointer to the next atom in the linked
  list for this hash entry;
\item \texttt{PropsOfAE}: a pointer to a linked list of atom \emph{properties};
\item \texttt{ARWLock}: a reader-writer lock that serializes access to the atom;
\item \texttt{StrOfAE}: a \texttt{C} representation of the atom's string.
\end{enumerate}

The \texttt{Prop} type abstracts objects that we refer to by the
atom's name. Example subtypes of \texttt{Prop} include functors,
modules, operators, global variables, blackboard entries, and
predicates. All of them are available by looking up an atom and
following the linked list of \texttt{Prop} objects.

Figure~\ref{fig:atom_table} shows an atom table with four atoms:
\texttt{hello}, \texttt{+}, \texttt{port}, and \verb+$live+. Notice
that only \texttt{+} and \verb+$live+ have associated properties. In
practice, most atoms do not have properties.

Every concrete type of  \texttt{Prop} implements two fields:
\begin{enumerate}
\item \texttt{NextOfPE} allows organizing properties for the same atom
  as a linked list;
\item \texttt{KindOfPE} gives the type of property.
\end{enumerate}

Each property extends the abstract property in its own way.
As an example, \emph{functors} add three extra fields: a back pointer
to the atom, the functor's arity, and a list of properties.

This design is based on LISP implementations, and  has been remarkably
stable throughout the history of the system.  Main optimizations and
extensions include: 

\begin{enumerate}
\item Recent versions of YAP support two atom tables: one groups all
  ISO-Latin-1 atoms, where each character code $c$ is such that $0 < c
  < 255$, and the other stores atoms that need to be represented as
  wide strings. YAP implements two tables in order to avoid an extra
  field per atom.
\item As discussed above, functors have their own \texttt{Prop}
  objects, namely, predicates and internal database keys with that
  functor.  This was implemented to improve performance of meta-calls.
\item The case where we have predicates with the same functor but
  belonging to different modules is addressed by a \emph{predicate
    hash-table}, which allows direct access to a predicate from a
  functor-module key.
\end{enumerate}

In Figure~\ref{fig:atom_table} the atom \texttt{+} has two properties:
one of the type \texttt{op} and another of type \texttt{functor}. The
atom \verb+$live+ has a property of type \texttt{predicate}.

Traditionally, YAP allocates memory as a single big chunk and then
uses its own memory allocator. This has two advantages: it avoids the
overhead of going through the standard \texttt{C} library, and it
simplifies state saving and restoring. The current allocator is Doug
Lea's global memory allocator~\cite{doug-lea-malloc}.

YAP can also use the \texttt{C}-library malloc,
as a configuration-time option. This is most useful in situations
where YAP needs to share memory with other systems (e.g, the Java
interface). 

\subsection{Dynamic Data Structures}

Each worker (parallel process or thread) maintains four separate
stacks and a set of registers. The stacks are:

\begin{itemize}
\item \emph{Global Stack}: implemented as an array of \texttt{CELL}s,
  it stores abstract objects of type \texttt{Term}.
\item \emph{Local Stack}: implemented as an array of \texttt{CELL}s,
  it stores choice-points and environments.
\item \emph{Trail}: it stores objects of type \texttt{TrailEntry}.
\item \emph{Auxiliary Stack} (generalizes the WAM PDL~\cite{Warren83}): a
  pointer to a reusable area of memory used to store temporary data,
  such as the stacks used for unification or other term matching
  operations.
\end{itemize}

Objects of type \texttt{Term} reside in the Global and Local Stacks,
and are always constructed from \texttt{CELL}s. YAP defines six
concrete types:

\begin{enumerate}
\item \emph{Small Integers}, are constructed from a subset of type
  \texttt{Int}, and occupy a single cell, with up to 29 bits in 32
  machines. \texttt{Int} is an integer--like type defined to take the
  same space as \texttt{CELL}.
\item \emph{Atoms}, occupy a single cell, and are constructed from
  objects of the abstract type \texttt{Atom}.
\item \emph{Applications}, or compound terms, occupy $N+1$ cells,
  where the first cell is reserved for an object of abstract type
  \texttt{Functor} with arity $N$, and the remaining $N$ cells for
  objects of type \texttt{Term}.
\item \emph{Pairs}, occupy $2$ cells, where the first cell is a
  \texttt{Term} called the \emph{Head}, and the second a \texttt{Term}
  called the \emph{Tail}.
\item \emph{References} occupy 1 cell, and are pointers to objects of
  type \texttt{Term}. By default, YAP represents free variables as
  self-references, but it can support free variables as the
  \texttt{NULL} pointer.
\item \emph{Extensions} occupy $N+2$ cells: the \emph{header}, a
  variable number $N$ of cells, and the \emph{footer}. The engine
  understands 3 extensions: floating point numbers, large integers,
  and \emph{blobs}, originally introduced to support very large
  integers.
\end{enumerate}

\subsection{Tagging Scheme}

Each different concrete type should have its own \emph{tag}. Tag
schemes differ significantly between Prolog
systems~\cite{schi90,DBLP:conf/plilp/TarauN94}; we refer the reader
to~\cite{DBLP:conf/ppdp/MoralesCH08} for a recent investigation of
this issue. The YAP tag scheme was designed to be
efficient~\cite{yap-optim}, and to allow using the whole available
memory in 32 bit machines. This allows at most $2$ bits for tags.

Unfortunately, there are six different concrete types: this
would require $\lceil log_2 6 \rceil = 3$ bits, but  in order to
access the full address space we are constrained to the $2$ lower
bits. The solution was:

\begin{itemize}
\item Atoms and Small Integers share the same tag. YAP allocates each
  atom as a separante object, guaranteeing that the object is always
  allocated at an address multiple of $8$, so that the third lowest
  bit can be used to distinguish between the two cases.
\item Applications and Extensions share the same tag. The header of an
  extension is a small number. This number is guaranteed to be an
  invalid address in modern systems, as these systems never allocate
  memory on the first virtual memory page.
\end{itemize}

This scheme allows for taking advantage of all the available memory
with a 32 bit \texttt{CELL}, but slows down access to compound terms.
A second drawback is that it requires explicit code for efficiency,
making it hard to take advantage of 64 bit machines.  Notice that YAP
does not need tag bits for garbage collection: instead, we use a
separate memory area to store the garbage collector
state~\cite{DBLP:journals/tplp/VandeginsteD07}.

Blobs were initially introduced to support very large numbers.  They enhance YAP
functionality without requiring extensive changes to the engine, and
currently provide the following functionality:

\begin{itemize}
\item \texttt{BIG\_INT}: very large integers, currently implemented
  through an interface to the GMP package~\cite{GMP}.
\item \texttt{BIG\_RATIONAL}: rationals.
\item \texttt{STRING}: sequences of character codes.
\item \texttt{EMPTY\_ARENA}: a chunk of cells that can be used to
  construct global variables or global data structures. This is used
  to support \texttt{nb\_} predicates and to implement
  \texttt{findall/3} and the \texttt{nb} library of global queues,
  global heaps, and global beams.
\item \texttt{ARRAY\_INT}: a multidimensional array of (non-tagged)
  integer numbers. This is manipulated by the package \texttt{matrix} .
\item \texttt{ARRAY\_FLOAT}: a multidimensional array of (non-tagged)
  floating-point numbers. This is manipulated by the package
  \texttt{matrix}.
\item \texttt{CLAUSE\_LIST}: a sequence of pointers to code. This
  allows for dynamic choice-points, and is used by the user-defined
  indexers~\cite{DBLP:conf/iclp/VazCF09}.
\end{itemize}

\section{The Engine}
\label{sec:engine}

YAP implements a version of David H.~D.~Warren's Warren Abstract
Machine (WAM)~\cite{Warren83}. Other Prolog systems using the WAM
include SICStus Prolog~\cite{SICS88}, Ciao Prolog~\cite{ciao},
XSB~\cite{Sagonas-98}, GNU Prolog~\cite{GNU-Prolog}, and
$ECL^iPS^e$~\cite{Eclipse}. The original machine consisted of $33$
instructions used to implement an environment-based term-copying
strategy. WAM instructions were divided into:

\begin{itemize}
\item Argument unification, or \texttt{get} and \texttt{unify}
  instructions; \item Argument building, or \texttt{put} instructions;
\item Control: \texttt{call}, \texttt{execute}, \texttt{proceed},
  \texttt{allocate}, and \texttt{deallocate};
\item Choice-point manipulation, or \texttt{try} instructions;
\item Indexing, or \texttt{switch} instructions;
\item Cut instructions.
\end{itemize}

The WAM instructions are very well suited to compilation: one compiles
a term by walking depth-first and left-to-right and associating each
symbol with an operation. Arguably, the WAM performs quite well and is
very well understood. On the other hand, most decisions on the WAM
were taken a long time ago, and there has been recent interest in
other abstract machine architectures for Prolog~\cite{DBLP:conf/iclp/Zhou07}.



YAP implements the YAAM emulator as a large \texttt{C} function. The
\texttt{C}-code for each instruction always starts with an \texttt{Op}
macro, and always terminates with an \texttt{EndOp} macro. Since YAP-6, a
Prolog program, \texttt{buildops}, understands these macros and uses
them to generate a file with all the YAAM opcodes, required by the
assembler, and a file with commands to restore a clause or to be
executed when the atom garbage collector needs to walk over YAAM
instructions. 

The emulator initializes by copying YAAM registers to local storage.
Whether this data is in the call stack or in the registers depends on
the Instruction Set Architecture, Operating Systems, Compiler, and
functionality being supported~\cite{yap-optim}. YAP then starts
executing YAAM instructions. Next, we discuss the main differences
between the YAAM~\cite{yap-optim} and the WAM.

\subsection{Unification Instructions}

There are several interesting issues regarding unification
instructions. A first problem is whether we should globalise void
variables occurring in the body of a clause. Consider the following
code fragment:

\begin{verbatim}
a(X) :- b(X,Y), c(X,Z).
\end{verbatim}

The variable \texttt{Y} is a void variable, and can be compiled either as a
\texttt{put\_x\_var} instruction, or as a \texttt{put\_y\_var}.
Compiling it as \texttt{put\_x\_var} requires placing the void
variable in the Global Stack: thus, space allocated to this variable
can only be recovered through backtracking, or through garbage
collection.  Compiling it as \texttt{put\_y\_var} requires placing the
variable in the Local Stack, and space can be recovered as soon as we
call \texttt{c/2}. We have experimented with both approaches, and
rarely noticed significant differences. YAP traditionally follows the
first approach, mostly in order to simplify compilation. Notice that
systems such as BIM-Prolog~\cite{BIM}, Aquarius~\cite{VanRoy90} or
hProlog~\cite{demoen00} address this problem simply by globalizing all
free variables..

A second problem is how to support nested
unification~\cite{MarienDSLP91}. Consider the clause

\begin{verbatim}
a([X,f(Y,X),Y]).
\end{verbatim}
\noindent
The WAM compiles the term breadth-first, obtaining:

\begin{verbatim}
a([X|Z]) :- Z = [A121|A122], A121 = f(Y,X), A122 = [Y].
\end{verbatim}
\noindent
Notice that the WAM approach requires extra temporary variables.
SICStus Prolog optimises the specific case of lists through the
\texttt{unify\_list} instruction that follow a list
depth-first~\cite{MatsCThesis}.

YAP implements a more general solution to this problem, first
published by Meier~\cite{NACLP90*63}. Sub-terms are always compiled
depth-first to \texttt{unify\_} instructions. Thus, YAP will generate
the following code in this case:

\begin{verbatim}
get_list         A1                  pop              1
unify_var        X1                  unify_last_list
unify_last_list                      unify_val        X2
unify_struct     f/2                 unify_last_atom  []
unify_var        X2                  proceed
unify_last_val   X1
\end{verbatim}

The code assumes an unification stack, initialized by the
\texttt{get\_list} and \texttt{get\_struct} instructions. Each
\texttt{unify\_list} or \texttt{unify\_struct} instruction pushes the
current state into the stack. The \texttt{pop} instruction pops this
state if necessary.

This algorithm is straightforward to implement; it does not put
pressure on temporary registers; and it allows inheritance of modes.
If YAP enters a structure in \texttt{write}-mode, then all its
sub-structures will execute in \texttt{write}-mode.

A corollary of this advantage is that \texttt{write} code performs
less tests, and is therefore simpler. This observation motivated the
double-opcode scheme originally presented in~\cite{yap-optim}. In this
method, each \texttt{unify} instruction has two opcodes, one taken in
\texttt{read} mode, and the other taken in \texttt{write} mode. It can
be easily shown that every \texttt{write-unify} instruction is
followed by:

\begin{itemize}
\item a \texttt{write-unify} instruction, if we are executing within
  the same sub-term;
\item a \texttt{pop} instruction, if we are moving to the parent sub-term;
\item a \texttt{get} or \texttt{control} instruction, if we have exited the
  term.
\end{itemize}

A similar argument can be made for most \texttt{read-unify}
instructions, with the exception of \texttt{unify\_list} and
\texttt{unify\_struct}. Therefore

\begin{itemize}
\item \texttt{unify\_list} and \texttt{unify\_struct} instructions
  either preserve \texttt{write} mode, or may move from
  \texttt{write} to \texttt{read} mode;
\item all other \texttt{unify\_} instructions preserve \texttt{write}
  mode;
\item \texttt{pop} instructions restore the \texttt{read/write} mode from
  the unification stack.
\end{itemize}

In our experience, separating write and read opcodes results in both a
faster and a cleaner engine implementation.

Other interesting design issues for compound terms include:

\begin{itemize}
\item In \texttt{get\_struct} and \texttt{unify\_struct} instructions,
  YAP immediately initializes the arguments of the newly generated
  compound term as unbound variables. YAP uses this eager approach
  because it be can be implemented as a tight loop, improving
  locality, and because it allows discarding \texttt{unify\_void}
  instructions at the end of a compound term.
\item YAP uses \texttt{unify\_last} instead of \texttt{unify} for the
  last instruction of a compound term. The \texttt{unify\_last}
  instructions do not need to update the \texttt{S} register,
  simplifying code. Moreover, \texttt{unify\_last\_list} and
  \texttt{unify\_last\_struct} instructions do not need to push the
  current state to the unification stack. 
\item YAP completes a sub-term when executing
  \texttt{unify\_last\_atom}, \texttt{unify\_last\_var} or
  \texttt{unify\_last\_val}. Next, YAP may need to execute a
  \texttt{pop} instruction to return to a sub-term above. The
  \texttt{pop} instruction will then set the read/write mode by
  choosing the read or write opcode of the next instruction.
\end{itemize}

\subsection{Term Sharing}

Consider the following clause:

\begin{verbatim}
a(X,W,[Y,Z]) :- b([Y,Z]), a(W,f([Y,Z])).
\end{verbatim}

The standard WAM would create three copies of \texttt{[Y,Z]}: one for
the head-term and two for the body-terms. YAP instead generates the
following code:

\begin{verbatim}
get_var           Y1,A2              allocate
get_list          A3                 call         user:b/1,1
save_pair         Y0                 put_val      Y1,A1
unify_var         X0                 put_struct   f/1,A2
unify_last_list                      write_val    Y0
unify_var         X0                 deallocate
unify_last_atom   []                 execute      user:a/2
put_val	          Y0,A1
\end{verbatim}

The \texttt{save\_pair} instruction stores $Abs Pair(\mathtt{S})$,
where \texttt{S} refers to the WAM's \texttt{S} structure pointer
register, in an abstract machine register or environment slot. This
argument is then stored in \texttt{Y0} in lieu of the term. This has two
advantages:

\begin{itemize}
\item Increased sharing, while reducing code size and run-time memory
  overhead;
\item Reduce the number of permanent variables. In the example,
  variables \texttt{Y} and \texttt{Z} are made void by this optimization. In
  contrast, the WAM would mark them as permanent variables.
\end{itemize}

The compiler maintains a table with all terms compiled so far in order
to support this operation. Every time a copy is found the term is
replaced by the argument. Notice that compilation times may increase
on large clauses with many terms. Thus, YAP thus imposes a maximum
threshold on the number of terms can be considered for reuse.

In a related optimization, recent versions of YAP compile \emph{large}
ground terms offline. That is, the clause

\begin{verbatim}
a("Long String").
\end{verbatim}

is compiled as:

\begin{verbatim}
get_dbterm  [76,111,110,103,32,83,116,114,105,110,103],A1
proceed
\end{verbatim}

Currently, \texttt{get\_dbterm} simply unifies its argument with a
ground term in the database. This has two advantages: it reduces code
size and it makes string construction constant-time. The major
drawbacks are the cost of maintaining an extra database of terms and
the need to implement JITI support.

\subsection{Non-logical Features}

Actual Prolog implementations must support non-logical features such
as the cut, disjunctions, and type predicates. YAP always stores a cut
pointer in the environment~\cite{Marien89:NACLP}. The implementation
of disjunction is more complex. Two basic approaches
are~\cite{MatsCThesis,disj-wks}:

\begin{itemize}
\item \emph{Offline} compilation~\cite{MatsCThesis} generates a new
  intermediate predicate and compiles disjuncts as new clauses. It
  allows for simpler compilation.
\item \emph{Inline} compilation uses special instructions to implement
  disjunction~\cite{disj-wks}.  It can reduce overheads.
\end{itemize}

YAP implements inline compilation of disjunctions. Each clause is
divided into a graph where an edge is an alternative to a disjunction,
and each edge starts with an \texttt{either}, an \texttt{or\_else}, or
\texttt{or\_last} instruction. These instructions implement a
choice-point with arity $0$, as all shared variables are guaranteed to
the environment.

As most other Prolog compilers, YAP also inlines a number of
built-ins~\cite{DBLP:conf/ppdp/NassenCS01,DBLP:conf/iclp/Zhou07}:

\begin{enumerate}
\item Type built-ins such as \texttt{var}, \texttt{nonvar},
  \texttt{atom} and related. They are implemented as \texttt{p\_var} ,
  \texttt{p\_nonvar} , \texttt{p\_nonvar} instructions.
\item Arithmetic operations. Currently, YAP only optimises integer
  operations. Examples include the \texttt{p\_plus} instructions,
  which are further optimised according to whether one of the
  arguments is a constant or not.
\item The \texttt{functor} and \texttt{arg} built-ins. YAP implements
  different \texttt{functor/3} instructions, depending on how
  arguments were instantiated at compile-time.
\item The meta-call: YAP inlines some
  meta-calls~\cite{DBLP:journals/tplp/TronconJDV07}. This is difficult,
  due to the complexity of the \texttt{goal\_expansion} and  the
  module mechanism.
\end{enumerate}
The implementation of inline built-ins has overgrown the initial
design, and requires redesign and a clean-up.

\section{Compilation}
\label{sec:compiler}

The YAP compiler 
implements the following steps algorithm:

\begin{enumerate}
\item \texttt{c\_head}: generate a WAM-like representation for the
  head of the clause.
\item If  the clause is a ground fact, proceed to step 6.
\item \texttt{c\_body}: generate WAM-like representation for the
  body of the clause.
\item \texttt{c\_layout}: perform variable classification and
  allocation.
\item \texttt{c\_optimize}: eliminate superfluous instructions.
\item Assemble the code and copy it to memory.
\end{enumerate}

The \texttt{c\_head} step simply walks over the clause head and
generates a sequence of WAM instructions.  The \texttt{c\_body}
routine visits the body goals and generates code for each goal in
sequence. Special care must be taken with disjunctions and with inline
built-ins.

Both \texttt{c\_head} and \texttt{c\_body} call \texttt{c\_goal}
to generate code for the head and sub-goals. The main challenge is to
compile variables, performed by \texttt{c\_var}. Each variable is
made to point to a \texttt{VarEntry} structure, that contains, among
other information: \textbf{(i)} a reference count indicating how many
times the variable was used in the clause; \textbf{(ii)} the first
occurrence of the variable in the code; and, \textbf{(iii)} the last
occurrence. The \texttt{c\_var} routine then works as follows:

\begin{itemize}
\item If this the first occurrence of the variable, \emph{bind} the
  variable to a \texttt{VarEntry}, set to have a reference count to
  $1$, and set the first and last occurrence to the current position.
\item Otherwise, increment reference count and set the
  last occurrence to the current position.
\end{itemize}
\texttt{c\_var} must also generate a WAM-like instruction for the
variable. It generates a \texttt{unify} instruction for variables in
sub-terms; a \texttt{put} instruction for variables in the body of the
clause; a \texttt{get} instruction for variables in the head.

The \texttt{c\_layout} routine proceeds as follows:

\begin{enumerate}
\item Reverse the chain of instructions.
\item Going from the end to the beginning, check if a variable must be
  permanent, and if so give it the next available environment
  slot. This guarantees that the environment variables occurring in
  the rightmost goals have the lower slots. This step again reverses
  the chain.
\item Going from the beginning to the end, allocate every temporary
  variable using a first-come, first-served greedy allocation
  algorithm. The YAAM has a very large array of registers, and spilling
  is considered an overflow.
\end{enumerate}

The \texttt{c\_optimize} step searches for unnecessary instructions,
say, \texttt{get\_x\_val A1,X1} and removes them.

\subsection{Compiling Disjunctions}

A clause with disjunctions can be understood as a directed acyclic
graph. Each node in the graph either delimits the beginning/end of the
clause or the beginning/end of a disjunction. Edges link nodes that
delimit an or-branch.  Notice that there is always an edge that
includes the head of the clause; we shall name this edge the
\emph{root-edge}.  Thus, a Horn Clause has two nodes and a single
edge, whereas a clause of the form

\begin{verbatim}
a :- (b ; c,d), e.
\end{verbatim}

has four nodes and four edges. YAP uses the
following principles to compile disjunctions:

\begin{itemize}
\item Any variable that crosses over two edges has to be
  initialized in the \emph{root-edge}. This prevents dangling
  variables, say:
\begin{verbatim}
g :- ( b(X) ; c(Y) ), d(Y).
\end{verbatim}
  \noindent The \texttt{Y} variable may be left dangling if not initialized
  before the edge.
\item As usual, environments are allocated if there is a path in the
  graph with two user-defined goals, or a user-defined goal followed
  by built-ins.
\item If a disjunction is of the form $G \rightarrow B_1 ; B_2$ and
  $G$ is a conjunction of \emph{test} built-ins, the compiler compiles
  $G$ with a jump to a \emph{fail} label that points to $B_2$.
\item Otherwise, the compiler generates choice-point manipulation
  instructions: the \texttt{either} instruction starts the
  disjunction; the \texttt{or\_else}  for inner edges; and the
  \texttt{or\_last} prepares the last edge for the disjunction.
\end{itemize}

There are cases where YAP has to do better. Consider a \emph{fast}
implementation of fibonacci:

\begin{verbatim}
fib(N, NX) :- ( N =< 1 ->
                NX = 1
              ;
                N1 is N - 1, N2 is N - 2,
                fib(N1, X1), fib(N2, X2),
                NX is X1 + X2
              ).
\end{verbatim}

The variables \texttt{N} and \texttt{NX} cross the disjunction,
therefore the above algorithm initializes them as permanent variables
at the root-edge. The problem is that the YAP variable allocator will
use the environment slots to access \texttt{N} and \texttt{NX}, and
would fail to take advantage of the fact that a \texttt{N} is
available in $A_1$ and \texttt{NX} in $A_2$. This generates
unnecessary accesses and the code may be less efficient than creating
a choice-point and executing a separate first clause. The solution is
to delay environment initialization until one is sure one needs
it. The rules are:

\begin{itemize}
\item Environments are allocated only \emph{once}: the edge that
  allocates the environment is the leftmost--topmost edge $E$ such
  that
  \begin{enumerate}
  \item no edge $E'$ above needs an environment, and,
  \item no edge to the left of $E$ needs the environment, and,
  \item $E$ or a descendant of $E$ needs the environment, and,
  \item at least a descendant of a right-sibling of $E$ needs the environment.
  \end{enumerate}
\item Variables are copied to the environment after allocation.
\end{itemize}

Applying these rules allows the compiler to delay marking some
variables as permanent variables. This simplifies the task of the
variable allocator, and leads to much faster code in the case above.

\subsection{The Assembler}
\label{sec:assembler}

The YAP Prolog assembler 
 converts from a high
level representation of YAAM instructions into YAAM byte-code. It
executes in two steps:

\begin{enumerate}
\item Compute addresses for labels and perform peephole optimizations,
  such as instruction merging.
\item Given the addresses of labels, copy instructions to actual
  location.
\end{enumerate}

Instruction
merging~\cite{yap-optim,demoen00,DBLP:conf/ppdp/NassenCS01,DBLP:conf/iclp/Zhou07}
is an important technique to improve performance of emulators. The
assembler implements instruction merging:

\begin{enumerate}
\item where it leads to improvement of performance in recursive
  predicates: examples include \texttt{get\_list} and
  \texttt{unify\_x\_val}, or \texttt{put\_y\_val} followed by
  \texttt{put\_y\_val}.
\item where it leads to substantial improvements in code size:
  examples include sequences of \texttt{get\_atom} instructions that
  are typical of database applications~\cite{DBLP:conf/padl/Costa07}.
\end{enumerate}

\section{The Just-In-Time Indexer}

\label{sec:jiti}

YAP includes a just-in-time indexer
(JITI)~\cite{jit-index,dyn-indexing}. Next, we give a brief overview
of how the algorithm has been implemented in the YAP system. First, we
observe that in YAP, in contrast to the WAM, by default predicates
have \emph{no} indexing code. All indexing is constructed at run-time.



Our first step is thus to ensure that calls to non-indexed predicates
have the abstract machine instruction \texttt{index\_pred} as their
first instruction.  This instruction calls the function
\texttt{Yap\_PredIsIndexable}, that implements the JITI.

\subsection{The Indexing Algorithm}

Indexing has been well studied in Prolog
systems~\cite{Carlsson,Bart89:NACLP,VanRoy90,ZhTaUs}. The main novelty in the
design of the JITI is that it tries to generate code that is
well-suited to the instantiations of the goal. To do so, it basically
follows a decision tree algorithm, where decisions are made by
inspecting the instantiation of the current call. The actual algorithm
is as follows:

\begin{enumerate}
\item Store pointers to every clause in the predicate in an array
  \texttt{Clauses} and compute the number of clauses.
\item Call \texttt{do\_index(Clauses,1)}, where the number $1$ refers
  to the first argument.
\item Assemble the generated code.
\end{enumerate}

The function \texttt{do\_index} is the core of the JITI. It is a
recursive function that, given a set of clauses $C$ with size $N$ and
an argument $i$, works as follows:

\begin{enumerate} 
\item If $N \leq 1$, call \texttt{do\_var} to handle the base case.
\item If $i > Arity$, we have tried every argument in the head: call
  \texttt{do\_var} to generate a
  \texttt{try}-\texttt{retry}-\texttt{trust} chain.
\item If $A_i$ is unbound, first call
  \texttt{suspend\_index(Clauses,i)}, to mark this argument
  as currently unindexed, and then call
  \texttt{do\_index(Clauses,i+1)}.
\item Extract the constraint that each clause $C$ imposes on $A_i$,
  and store the constraint in $Clauses[C]$. The YAP JITI understands
  two types of constraints:
  \begin{itemize}
  \item bindings, of the form $X = T$, where the main functor of $T$
    is known;
  \item type-constraints, such as $number(X)$.
  \end{itemize}
\item Compute the \emph{groups}, where a group is a contiguous subset
  of clauses that can be  indexed through a single \texttt{switch\_on\_type}~\cite{Warren83}. For example, consider the following
  definition of predicate \texttt{a/1}:

\begin{verbatim}
a(1). a(1). a(2). a(X). a(1).
\end{verbatim}    

  This predicate has three groups: the first three clauses form a
  group, and the fourth and fifth clauses form each one a different
  group. The fourth clause forms a \emph{free} group, as it
  imposes no constraint on $A_1$.
\item In order to generate simpler code, if the number of groups $NG$,
  is larger than one and we are not looking at the first argument,
  that is $ NG > 1 \land i > 1$, then do not try indexing the current
  argument, and instead call \texttt{do\_index(Clauses,i+1)}.
\item Compile the groups one by one.  If the group is free, call
  \texttt{do\_var}: this function generates the leaf code for a
  sequence of 
  \texttt{try}-\texttt{retry}-\texttt{trust} instructions.

  Otherwise, if all constraints in the group are binding constraints:
  \begin{enumerate}
  \item generate a \texttt{switch\_on\_type} instruction for the
    current argument $i$;
  \item The \texttt{switch\_on\_type} instruction has 4 slots in the
    YAAM (and in the WAM): constants, compound terms, pairs, and
    unbound variables. The JITI generates code for the first three
    cases. The fourth case is not compiled for; instead the JITI fills
    the last slot with the \emph{expand\_index} instruction (discussed
    in detail later).
  \item Next, separate clauses in three subgroups according to whether
    they contain a constant (atoms or small integers), a pair, or a
    compound term, including extensions.
  \item Call \texttt{do\_consts}, \texttt{do\_funcs}, and
    \texttt{do\_pair} on each subgroup to fill in the remaining slots.
  \end{enumerate}

A clause imposing a type-constraint requires specialized processing, for example:
\begin{enumerate}
  \item $integer(A_i)$ adds the clause to the list of constants and to
    the list of functors.
  \item $var(A_i)$ requires removing the current clause from the list
    of constants, functors and pairs;
  \item $nonvar(A_i)$ cannot select between different cases, and is
    not used.
  \end{enumerate}
\end{enumerate}

The \texttt{do\_var} auxiliary routine is called to handle cases
where we cannot index further: it either commits to a clause, or
creates a chain of \texttt{try}-\texttt{retry}-\texttt{trust}
instructions. The \texttt{do\_consts}, \texttt{do\_funcs}, and
\texttt{do\_pair} functions try to construct a decision list or hash
table on the values of the main functor of the current term, in a
fashion very similar to the standard WAM. On the other hand,
\texttt{do\_funcs}, and \texttt{do\_pair} may call
\texttt{do\_compound\_index} to index on sub-terms. Finally, YAP
implements a few optimizations to handle common cases that do not fit
well in this algorithm (e.g., catch-all clauses).

The \texttt{suspend\_index(Clauses,i)} function generates an
\texttt{expand\_index $A_i$} instruction at the current location, and
then continues to the next argument. At run time, if ever the
instruction is visited with $A_i$ bound, YAP will expand the index
tree, as discussed next.

\subsection{Expanding The Index Tree}
\label{sec:expanding-index-tree}

The \texttt{expand\_index} YAAM instruction verifies whether new calls to
the indexing code have the same instantiation as the original
call. Thus, it allows the YAP JITI to grow the tree whenever we
receive calls with different modes. The instruction executes as
follows. First, it recovers
the $PredEntry$ for the current predicate, and then it calls
\texttt{Yap\_ExpandIndex} that proceeds as follows:

\begin{enumerate}
\item Initialize clause and groups information.
\item Walk the indexing tree from scratch, finding out which
  instruction caused the transfer to \texttt{expand\_index}, and what
  clauses matched at that point. Store the matching clauses in the
  $Clauses$ array.
\item Call \texttt{do\_index(Clauses, i+1)} to construct the new tree;
\item Link the new tree back to the current indexing tree.
\end{enumerate}

The second step is required because when we call
\texttt{expand\_index} we do not actually have a pointer to the
previous instruction, nor do we know how many clauses do match at this
point (doing so would very much increase the size of the indexing
code). Instead, we have to follow the indexing code again from
scratch. As \texttt{Yap\_ExpandIndex} executes each instruction in
the indexing tree, it also selects the clauses that can still
match. The algorithm is as follows:

\begin{enumerate}
\item Set the alternative program pointer, $AP$ to \texttt{NULL}, the
  parent program pointer $P'$ to \texttt{NULL}, and the program
  pointer $P$ to point at the initial indexing instruction.
\item While the YAAM instruction \texttt{expand\_index} was not found:
\item Set the current instruction pointer $P$ to  be $P'$.
\item If the current opcode is:
  \begin{itemize}
  \item \texttt{switch\_on\_type} then check the type of the current
    argument $i$, remove all clauses that are constrained to a
    different type from $Clauses$, and compute the new $P$.
  \item \texttt{switch\_on\_\{cons, struct\}} then check if the
    current argument $i$ matches one of the constants (functors). If
    so, remove all clauses that are constrained to a different
    constant from $Clauses$, and take the corresponding entry. If not,
    jump to $AP$.
  \item \texttt{try} then mark that we are not the first clause and set
    $AP$ to the next instruction.
  \item \texttt{retry} then set $AP$ to be the next instruction and
    jump to the label.
  \item \texttt{trust} set then $AP$ to \texttt{NULL} and jump to the
    label.
  \item \texttt{jump\_if\_nonvar} then check if the current $A_i$ is
    bound. If not, proceed to the next instruction. Otherwise, if the
    jump label is \texttt{expand\_index}, we are done.
  \end{itemize}
\end{enumerate}

The algorithm returns a set of clauses $Clauses$ and a pointer $P'$
giving where the code was called from.  We thus can call
\texttt{do\_index} as if it had been called from the
\texttt{index\_pred} instruction.

\subsection{The JITI: Discussion}
\label{sec:jiti-discussion}
The main advantages of the JITI are the ability to index multiple
arguments and compound terms, and the ability to index for multiple
modes of usage. Several Prolog systems do support indexing on multiple
arguments~\cite{SWI,DBLP:conf/iclp/Zhou07,xsb-manual}; on the other
hand, we are not aware of other systems that allow multiple modes. Our
experience has shown that this feature is very useful in applications
with large databases. A typical example is where we use the database
to represent a graph and we want to walk edges in \emph{both}
directions; a second typical application is when mining
databases~\cite{fonseca-efficiency}. Arguably, a smart programmer will
be able to address these problems by duplicating the database: the
JITI is about not having to do the effort.

The JITI has a cost. First, the index size can grow significantly, and
in fact exceed the size of the original
database~\cite{fonseca-efficiency}. In the worst case we can build a
large index that will serve a single call. Fortunately, our experience
has shown this to be rare. In most cases, if the index grows, it is
because it is needed, and the benefits in running-time outweigh the
cost in memory space. A second drawback is the cost of calling
\texttt{Yap\_ExpandIndex}. Although we have not systematically measured
this overhead, in our experience it is small.

\section{OPTYAP: An Overview}
\label{sec:optyap}

One of the major advantages of Logic Programming is that it is well
suited for parallel execution. The interest in the parallel execution
of logic programs mainly arose from the fact that parallelism can be
exploited \emph{implicitly}, that is, without input from the
programmer to express or manage parallelism, ideally making Parallel
Logic Programming as easy as Logic Programming.

On the other hand, the good results obtained with
tabling~\cite{xsb-manual} raise the question of whether further
efficiency would be achievable through parallelism. Ideally, we would
like to exploit maximum parallelism and take maximum advantage of
current technology for tabling and parallel systems. Towards this
goal, we proposed the \emph{Or-Parallelism within Tabling}
(\emph{OPT}) model. The OPT model generalizes Warren's
multi-sequential engine framework for the exploitation of
or-parallelism in shared-memory models. It is based on the idea that
all open alternatives in the search tree should be amenable to
parallel exploitation, be they from tabled or non-tabled subgoals.
Further, the OPT model assumes that tabling is the base component of
the parallel system, that is, each \emph{worker} is a full sequential
tabling engine, the or-parallel component only being triggered when
workers run out of alternatives to exploit.

OPTYAP implements the OPT model, and we shall use the name OPTYAP to
refer to YAP plus tabling and
or-parallelism~\cite{optyap-journal}. OPTYAP builds on the
YAPOR~\cite{Rocha-99b} and YAPTAB~\cite{Rocha-00a} work. YAPOR was
previous work on supporting or-parallelism over
YAP~\cite{Rocha-99b}. YAPOR is based on the environment copying model
for shared-memory machines, as originally implemented in
Muse~\cite{Ali-90b}. YAPTAB is a sequential tabling engine that
extends YAP's execution model to support tabled evaluation for
definite programs. YAPTAB's implementation is largely based on the
seminal design of the XSB system, the SLG-WAM~\cite{Sagonas-98}, but
it was designed for eventual integration with YAPOR. Parallel tabling
with OPTYAP is implicitly triggered when both YAPOR and YAPTAB are
enabled.

\subsection{The Sequential Tabling Engine}

Tabling is about storing intermediate answers for subgoals so that
they can be reused when a \emph{variant call}\footnote{Two calls are
  said to be variants if they are the same up to variable renaming.}
appears during the resolution process. Whenever a tabled subgoal is
first called, a new entry is allocated in an appropriate data space,
the \emph{table space}. Table entries are used to collect the answers
found for their corresponding subgoals. Moreover, they are also used
to verify whether calls to subgoals are variant. Variant calls to
tabled subgoals are not re-evaluated against the program clauses,
instead they are resolved by consuming the answers already stored in
their table entries. During this process, as further new answers are
found, they are stored in their tables and later returned to all
variant calls. Within this model, the nodes in the search space are
classified as either: \emph{generator nodes}, corresponding to first
calls to tabled subgoals; \emph{consumer nodes}, corresponding to
variant calls to tabled subgoals; or \emph{interior nodes},
corresponding to non-tabled subgoals.

To support tabling, YAPTAB introduces a new data area to the YAP
engine, the \emph{table space}, implemented using
\emph{tries}~\cite{RamakrishnanIV-99}; a new set of registers, the
\emph{freeze registers}; an extension of the standard trail, the
\emph{forward trail}; and four new operations for tabling. The
configuration macro \texttt{TABLING} defines when tabling support is
enabled in YAP. 
The new tabling operations are:

\begin{itemize}
\item The \emph{tabled subgoal call} operation checks if a subgoal is
  a variant call. If so, it allocates a consumer node and starts
  consuming the available answers. If not, it allocates a generator
  node and adds a new entry to the table space. Generator and consumer
  nodes are implemented as standard choice points extended with an
  extra field, \texttt{cp\_dep\_fr}, that is a pointer to a
  \emph{dependency frame} data structure used by the \emph{completion}
  procedure. Generator choice points include another extra field,
  \texttt{cp\_sg\_fr}, that is a pointer to the associated
  \emph{subgoal frame} where tabled answers should be stored. Tabled
  predicates defined by several clauses are compiled using the
  \texttt{table\_try\_me}, \texttt{table\_retry\_me} and
  \texttt{table\_trust\_me} WAM-like instructions in a manner similar
  to the generic
  \texttt{try\_me}/\texttt{retry\_me}/\texttt{trust\_me} WAM
  sequence. The \texttt{table\_try\_me} instruction extends the WAM's
  \texttt{try\_me} instruction to support the tabled subgoal call
  operation. The \texttt{table\_retry\_me} and
  \texttt{table\_trust\_me} differ from the generic WAM instructions
  in that they restore a generator choice point rather than a standard
  WAM choice point. Tabled predicates defined by a single clause are
  compiled using the \texttt{table\_try\_single} WAM-like instruction,
  a specialized version of the \texttt{table\_try\_me} instruction for
  deterministic tabled calls.
\item The \emph{new answer} operation checks whether a newly found
  answer is already in the table, and if not, inserts the
  answer. Otherwise, the operation fails. The
  \texttt{table\_new\_answer} instruction implements this operation.
\item The \emph{answer resolution} operation checks whether extra
  answers are available for a particular consumer node and, if so,
  consumes the next one. If no answers are available, it suspends the
  current computation and schedules a possible resolution to continue
  the execution. It is implemented by the
  \texttt{table\_answer\_resolution} instruction.
\item The \emph{completion} operation determines whether a subgoal is
  completely evaluated and when this is the case, it closes the
  subgoal's table entry and reclaims stack space. Otherwise, control
  moves to one of the consumers with unconsumed answers. The
  \texttt{table\_completion} instruction implements it. On completion
  of a subgoal, the strategy to implement answer retrieval consists in
  a top-down traversal of the completed answer tries and in executing
  dynamically compiled WAM-like instructions from the answer trie
  nodes. These dynamically compiled instructions are called \emph{trie
    instructions} and the answer tries that consist of these
  instructions are called \emph{compiled
    tries}~\cite{RamakrishnanIV-99}. 
\end{itemize}

Completion is hard because a number of subgoals may be mutually
dependent, thus forming a \emph{Strongly Connected Component} (or
\emph{SCC})~\cite{Tarjan-72}. The subgoals in an SCC are completed
together when backtracking to  the \emph{leader node} for the SCC, i.e., the
youngest generator node that does not depend on older
generators. YAPTAB innovates by considering that the control of
completion detection and scheduling of unconsumed answers should be
performed through the data structures corresponding to variant calls
to tabled subgoals, and does so by associating a new data structure,
the \emph{dependency frame}, to consumer nodes. Dependency frames are
used to efficiently check for completion points and to efficiently
move across the consumer nodes with unconsumed answers. Moreover, they
allow us to eliminate the need for a separate completion stack, as
used in SLG-WAM's design, and to reduce the number of extra fields in
tabled choice points. Dependency frames are also the key
data structure to support parallel tabling in OPTYAP.

Another original aspect of the YAPTAB design is its support for the
dynamic mixed-strategy evaluation of tabled logic programs using
batched and local scheduling~\cite{Rocha-05c}, that is, it allows one
to modify at run-time the strategy to be used to resolve the
subsequent subgoal calls of a tabled predicate. At the engine level,
this includes minor changes to the tabled subgoal call, new answer and
completion operations, all the other tabling extensions being commonly
used across both strategies.

More recent contributions to YAPTAB's design include the proposals to
efficiently handle incomplete and complete
tables~\cite{Rocha-06a}. Incomplete tables are a problem when, as a
result of a pruning operation, the computational state of a tabled
subgoal is removed from the execution stacks before being
completed. In such cases, we cannot trust the answers from an
incomplete table because we may loose part of the computation. YAPTAB
implements an approach where it keeps incomplete tables around and
whenever a new variant call for an incomplete table appears, it first
consumes the available answers and only if the table is exhausted, it
will restart the evaluation from the beginning. This approach avoids
re-computation when the already stored answers are enough to evaluate
the variant call. On the other hand, complete tables can also be a
problem when we use tabling for applications that build arbitrarily
many large tables, quickly exhausting memory space. In general, we
will have no choice but to throw away some of the tables (ideally, the
least likely to be used next). YAPTAB implements a memory management
strategy based on a \emph{least recently used} algorithm for the
tables. With this approach, the programmer can rely on the
effectiveness of the memory management algorithm to completely avoid
the problem of deciding what potentially useless tables should be
deleted.

Performance results for YAPTAB have been very encouraging from the
beginning. Initial results showed that, on average, YAPTAB introduces
an overhead of about 5\% over standard Yap when executing non-tabled
programs~\cite{Rocha-00a}. For tabled programs, results indicated that
we successfully accomplished our initial goal of comparing favorably
with current state-of-the-art technology since, on average, YAPTAB
showed to be about twice as fast as XSB~\cite{Rocha-00a}. In more
recent studies, comparing YAPTAB with other tabling Prolog systems,
the previous results were confirmed and YAPTAB showed to be, on
average, twice as fast as XSB and Mercury~\cite{Somogyi-06} and more
than twice faster than Ciao Prolog and
B-Prolog~\cite{Chico-08}. Regarding the overhead for supporting
mixed-strategy evaluation, our results showed that, on average, YAPTAB
is about 1\% slower when compared with YAPTAB supporting a single
scheduling strategy~\cite{Rocha-05c}. Moreover, our results showed
that dynamic mixed-strategies, incomplete tabling and table memory
recovery can be extremely important to improve the performance and
increase the size of the problems that can be solved for ILP-like
applications~\cite{Rocha-07a}. Considering that YAP is one of the
fastest Prolog engines currently available, these results are quite
satisfactory and they show that YAPTAB is a very competitive tabling
system.

\subsection{The Or-Parallel Tabling Engine}


In OPTYAP, or-parallelism is implemented through copying of the
execution stacks. More precisely, we optimize copying by using
\emph{incremental copying}, where workers only copy the differences
between their stacks. All other YAP areas and the table space are
shared between workers. Incremental copying is part of YAPOR's engine.

A first problem that we had to address in OPTYAP was concurrent access
to the table space. OPTYAP implements four alternative locking schemes
to deal with concurrent accesses to the table space data structures,
the \emph{Table Lock at Entry Level} (TLEL) scheme, the \emph{Table
  Lock at Node Level} (TLNL) scheme, the \emph{Table Lock at Write
  Level} (TLWL) scheme, and the \emph{Table Lock at Write Level -
  Allocate Before Check} (TLWL-ABC) scheme. The TLEL scheme includes a
single lock per trie, and thus allows a single writer per trie. The
TLNL has a lock per node, and thus allows a single worker per chain of
sibling nodes that represent alternative paths from a common parent
node. The TLWL scheme is similar to TLNL but the common parent node is
only locked when writing to the table is likely. Lastly, the TLWL-ABC is
an optimization that allocates and initializates nodes that are likely
to be inserted in the table space before any locking is
performed. Experimental results~\cite{rocha_ippdps02} showed that TLWL
and TLWL-ABC present the best speedup ratios and that they are the
only schemes showing good scalability.

A second problem was \emph{public completion}. When a worker $W$
reaches a leader node for an SCC $S$ and the node is public, other
workers can still influence $S$, for example, if finding new answers
for consumers in $S$. In such cases, $W$ cannot complete but, on the
other hand, it would like to move elsewhere in the tree to try other
work. Note that this is the only case where or-parallelism and tabling
conflict. One solution would be to disallow movement in this case.
Unfortunately, we would severely restrict parallelism. As a result, in
order to allow $W$ to continue execution it becomes necessary to
\emph{suspend the SCC} at hand. Suspending an SCC consists of saving
the SCC's stacks to a proper space and leave in the leader node a
reference to the suspended SCC. These suspended computations are
reconsidered when the remaining workers perform the completion
operation. Thus, an SCC $S$ is completely evaluated when the
following two conditions hold:

\begin{itemize}
\item There are no unconsumed answers in any consumer node belonging
  to $S$ or in any consumer node within a suspended SCC in a node
  belonging to $S$.
\item There are no other representations of the leader node $L$ in the
  computational environment. In other words, $L$ cannot be found in
  the execution stacksf of a different worker, and $L$ cannot be found
  in the suspended stack segments for another SCC.
\end{itemize}

Knowing that worker $W$ is at
 the current leader node $L$ for an SCC
$S$, the algorithm for public completion is actually quite
straightforward:

\begin{itemize}
\item \emph{Atomically} check whether $W$ is the last worker at node
  $L$, and remember the result as a boolean variable
  \texttt{LastWorkerAtNode}.
\item Check if there are unconsumed answers in any consumer node
  belonging to $S$ or in any consumer node within a suspended SCC in a
  node belonging to $S$. If so, resume and move to this work.
\item If \texttt{LastWorkerAtNode} is false, suspend the current SCC and
  call the \emph{scheduler} to get a new piece of unexploited work.
\item Otherwise, if \texttt{LastWorkerAtNode} is true, $W$ has
  completed.
\end{itemize}

The synchronization corresponds to checking beforehand whether $W$ is
the last worker, and if so, complete. Note that $W$'s code must take
care to check whether $W$ is last before it checks for uncompleted
answers, as new answers or nodes might have been generated
meanwhile. 

A worker $W$ enters in scheduling mode when it runs out of work and
only returns to execution mode when a new piece of unexploited work is
assigned to it by the scheduler. The scheduler must efficiently
distribute the available work for exploitation between workers.
OPTYAP has the extra constraint of keeping the correctness of
sequential tabling semantics. The OPTYAP scheduler is essentially the YAPOR
scheduler~\cite{Rocha-99b}: \emph{when a worker runs out of work it
  searches for the nearest unexploited alternative in its branch. If
  there is no such alternative, it selects a busy worker with excess
  of work load to share work with. If there is no such a worker, the
  idle worker tries to move to a better position in the search
  tree}. However, some extensions were introduced in order to preserve
the correctness of tabling semantics and to ensure that a worker never
moves above a leader until it has fully exploited all
alternatives. Thus, OPTYAP introduces the constraint that the
\emph{computation cannot flow outside the current SCC, and workers
  cannot be scheduled to execute at nodes older than their current
  leader node}.

Parallel execution of tabled programs in OPTYAP showed that the system
was able to achieve excellent speedups up to 32 workers for
applications with coarse grained parallelism and quite good speedups
for applications with medium parallelism~\cite{optyap-journal}.

\section{Future Challenges}
\label{sec:challenges}

Prolog is a well-known language. It is widely used, and it is a
remarkably powerful tool. The core of Prolog has been very stable
throughout the years, both in terms of language design and in terms of
implementation. Yet, there have been several developments, many within
the Logic Programming community, and many more outside.  Addressing
these developments and the needs of a world very different from when
Prolog was created, presents both difficulties and opportunities.
Next, we discuss some of these issues from our personal perspective.


\paragraph{Compiler Implementation Technology} 

Implementation technology in Prolog needs to be rethought. At the
low-level, only GNU Prolog currently generates
native-code~\cite{GNU-Prolog}. Just-In-Time technology is a natural
match to Prolog and it has shown to work well, but we have just
scratched the surface~\cite{DBLP:conf/iclp/SilvaC07}. Progress in
compilers, such as \texttt{GCC}, may make compilation to \texttt{C}
affordable again. At a higher level, more compile-time optimization
should be done. Determinacy detection is well
known~\cite{Dawson:1995:UFE} and should be available. Simple
techniques, such as query reordering, can change program performance
hugely for database queries. They should be easily available.

A step further: code expansion for recursive procedures is less of a
problem, so why not rethink old ideas such as Krall's
VAM~\cite{Krall96}, and Beer's uninitialized
variables~\cite{Beer,VaRo92}?  Moreover, years of experience with Ciao
Prolog should provide a good basis for rethinking global
analysis~\cite{ciao-effective}.

Last, but not least, Prolog implementation is not just about pure Horn
clauses. Challenges such as negation~\cite{xsb-manual} and
coinduction~\cite{DBLP:conf/iclp/SimonMBG06} loom large over the
future of Logic Programming.

\paragraph{Language Technology} 

At this point in time, there is no dominant language nor
framework. But, arguably, some lessons can be taken:
\begin{itemize}
\item \emph{Libraries and Data-Structures}: languages need to provide useful,
  reusable code.
\item \emph{Interfacing}: it should be easy to communicate with other
  languages, and especially with domain specific languages, such as
  \texttt{SQL} for databases, and \texttt{R} for statistics.
\item \emph{Typing}: it is not clear whether static typing is needed,
  but it is clear that it is useful, and that it is popular in the
  research community.
\end{itemize}

Our belief is that progress in this area requires collaboration
between different Prolog systems, namely so that it will be easy to
reuse libraries and code. YAP and SWI-Prolog are working together in
this direction.

\paragraph{Logic Programming Technology} 

Experience has shown that it is hard to move results from Logic
Programming research to Prolog systems. One illustrative example is
XSB Prolog~\cite{xsb-manual}: on the one hand, the XSB system has been
a vehicle for progress in Logic Programming, supporting the tabling of
definite and normal programs. On the other hand, progress in XSB
has not been widely adopted. After more 10 years, even tabling of
definite programs is not widely available in other Prolog systems.

The main reason for that is complexity: it is just very hard to
implement some of the novel ideas proposed in Logic Programming. Old
work suggests that Logic Programming itself may help in this
direction~\cite{SLGresolution}. Making it easy to change and control
Prolog execution in a flexible way is a fundamental challenge for
Prolog.
 
\paragraph{The WWW} 

It has become very important to be able to reason and manipulate data
on the world-wide web. Surprisingly, one can see relatively little
contribution from the Logic Programming community, although it should
be clear that Prolog can have a major role to play, especially related
to the semantic web~\cite{DBLP:conf/semweb/WielemakerHOS08}. Initial
results offer hope that YAPTAB is competitive with specialized systems
in this area~\cite{DBLP:conf/www/LiangFWK09}.

\paragraph{Uncertainty}

The last few years have seen much interest in what is often called
Statistical Relational Learning (SRL). Several languages designed for
this purpose build directly upon Prolog. PRISM~\cite{prism} is one of
the most popular examples: progress in PRISM has stimulated progress
in the underlying Prolog system, B-Prolog~\cite{DBLP:conf/iclp/Zhou07}. Problog is an exciting
recent development, and supporting Problog has already lead to
progress in YAP~\cite{DBLP:conf/iclp/KimmigCRDR08}.

Note that even SRL languages that \emph{do not} rely on Prolog offer
interesting challenges to the Prolog community. As an interesting
example, Markov Logic Networks (MLNs)~\cite{RiMo06} are a popular SRL
language that uses bottom-up inference and incremental query
evaluation, two techniques that have been well researched in Logic
Programming.

\section{Conclusions and Future Work}
\label{sec:conclusions}

We presented the YAP system, gave the main principles of its
implementation, and detailed what we believe are the main
contributions in the design of the system, such as engine design,
just-in-time-indexing, tabling, and parallelism. Arguably, these
contributions have made YAP a very competitive system in Prolog
applications that require access to large amounts of data, such as
learning applications.

Our experience, both as implementers and as users, shows that there
are a number of challenges to Prolog. We would like to make ``Prolog''
faster, more attractive to the Computer Science community and, above all, more
useful. To do so, much work has still to be done. Some of the
immediate work ahead includes integrating the just-in-time clause
compilation framework in the main design of the system, improving
performance for attributed variables and constraint systems, improving
compatibility with other Prolog systems, and, as always, fixing bugs.

We discussed some of the main challenges that in our opinion face
Logic Programming above. YAP has also shown to be an useful platform
for work in the languages that combine Prolog and probabilistic
reasoning, such as CLP($\cal BN$)~\cite{DBLP:conf/ilp/CostaPC08},
ProbLog~\cite{DBLP:conf/iclp/KimmigCRDR08},
and CPlint~\cite{DBLP:conf/aiia/Riguzzi07}.  As argued above, we
believe this is an important research direction for the Logic
Programming community, and plan to pursue this work further.

\subsection*{Acknowledgments}
YAP would not exist without the support of the YAP users. We would
like to thank them first.  The work presented in this
paper has been partially supported by project HORUS
(PTDC/EIA-EIA/100897/2008), LEAP (PTDC/EIA-CCO/112158/2009), and funds
granted to {\em LIACC} and \emph{CRACS \& INESC-Porto LA} through the
{\em Programa de Financiamento Plurianual, Funda\c{c}\~ao para a
  Ci\^encia e Tecnologia} and {\em Programa POSI}. Last, but not
least, we would like to gratefully acknowledge the anonymous referees
and the editors of the special number for the major contributions that
they have given to this paper.

\bibliography{lp}

\end{document}